\documentclass{article}%
\usepackage{amssymb}
\usepackage{amsmath}%
\usepackage{amsfonts}%
\usepackage{graphicx}
\providecommand{\U}[1]{\protect\rule{.1in}{.1in}}
\newtheorem{theorem}{Theorem}

\newtheorem{definition}[theorem]{Definition}

\newtheorem{lemma}[theorem]{Lemma}

\newtheorem{remark}[theorem]{Remark}

\begin{document}

\title{On the existence of stable charged Q-balls.}
\author{Vieri Benci$^{\ast}$, Donato Fortunato$^{\ast\ast}$\\$^{\ast}$Dipartimento di Matematica Applicata \textquotedblleft U.
Dini\textquotedblright\\Universit\`{a} degli Studi di Pisa \\Largo Bruno Pontecorvo 1/c, 56127 Pisa, Italy\\e-mail: benci@dma.unipi.it\\$^{\ast\ast}$Dipartimento di Matematica \\Universit\`{a} degli Studi di Bari Aldo Moro\\Via Orabona 4, 70125 Bari, Italy\\e-mail: fortunat@dm.uniba.it}
\maketitle

\begin{abstract}
This paper concerns hylomorphic solitons, namely stable, solitary waves whose
existence is related to the ratio energy/charge. In theoretical physics, the
name \textit{Q-ball} refers to a type of hylomorphic solitons or solitary
waves relative to the Nonlinear Klein-Gordon equation (NKG). We are interested
in the existence of charged Q-balls, namely Q-balls for the Nonlinear
Klein-Gordon equation coupled with the Maxwell equations (NKGM). In this case
the charge reduces to the electric charge. The main result of this paper
establishes that stable, charged Q-balls exist provided that the interaction
between matter and the gauge field is sufficiently small.

\end{abstract}
\tableofcontents

\bigskip


\bigskip


\section{Introduction}

Roughly speaking a \textit{solitary wave} is a solution of a field equation
whose energy travels as a localized packet and which preserves this
localization in time. A \textit{soliton} is a solitary wave which exhibits
some form of stability so that it has a particle-like behavior (see e.g.
\cite{Ba-Be-R.}, \cite{befogranas}, \cite{raj}, \cite{vil}).\ 

Following \cite{milano}, a soliton or solitary wave is called
\textit{hylomorphic} if its stability is due to a particular ratio between
\textit{energy} and the \textit{hylenic} \textit{charge}. The hylenic charge
is an integral of motion due to a $S^{1}$-invariance of the Lagrangian (cf.
section \ref{im} and for more details see \cite{milano}).

In theoretical physics, the name \textit{Q-ball} refers to a type of
hylomorphic solitons or solitary waves relative to the Nonlinear Klein-Gordon
equation (NKG). This type of solitary waves has been first studied in the
pioneering paper \cite{rosen68}. The name Q-ball has been introduced by
Coleman in \cite{Coleman86}. We recall that stability results for (NKG) has
been established in \cite{BBBM}, \cite{hylo}, \cite{befon}, \cite{bonanno},
\cite{shatah}. If the Klein-Gordon equations are coupled with the Maxwell
equations (NKGM) then the hylenic charge reduces to the electric charge and
the relative solitary waves are called \textit{charged}, or \textit{gauged
Q-balls} (see e.g.\cite{Coleman86}). Existence results of Q-balls for (NKGM)
are stated in, \cite{befo}, \cite{befo08}, \cite{befo10}, \cite{mug}, however,
in these papers there are not stability results and hence the existence of
solitons for (NKGM) (namely \textit{stable} charged Q-balls) was an open question.

The problem with the stability of charged Q-ball is that the electric charge
tends to brake them since charges of the same sign repel each other. In this
respect Coleman in his celebrated paper \cite{Coleman86} says "\textit{I have
been unable to construct Q-balls when the continuous symmetry is gauged. I
think what is happening physically is that the long-range force caused by the
gauge field forces the charge inside the Q-ball to migrate to the surface, and
this destabilizes the system, but I am not sure of this}".

The aim of this paper is to investigate this problem. Our main result
establishes that stable charged Q-balls exist provided that the interaction
between matter and gauge field is sufficiently small. Thus, this paper gives a
partial answer to the problem risen by Coleman. In order to give a complete
answer, we should know what happens if the interaction is strong. We
conjecture that, in this case, charged stable Q-balls do not exist, but now we
do not have yet a complete proof.

\section{Hylomorphic solitary waves and solitons\label{HS}}

\subsection{An abstract definition of solitary waves and solitons\label{be}}

Solitary waves and solitons are particular \textit{states} of a dynamical
system described by one or more partial differential equations. Thus, we
assume that the states of this system are described by one or more
\textit{fields} which mathematically are represented by functions
\begin{equation}
\mathbf{u}:\mathbb{R}^{N}\rightarrow V\label{lilla}%
\end{equation}
where $V$ is a vector space with norm $\left\vert \ \cdot\ \right\vert _{V}$
which is called the internal parameters space. We assume the system to be
deterministic; this means that it can be described as a dynamical system
$\left(  X,\gamma\right)  $ where $X$ is the set of the states and
$\gamma:\mathbb{R}\times X\rightarrow X$ is the time evolution map. If
$\mathbf{u}_{0}(x)\in X,$ the evolution of the system will be described by the
function
\begin{equation}
\mathbf{u}\left(  t,x\right)  :=\gamma_{t}\mathbf{u}_{0}(x).\label{flusso}%
\end{equation}
We assume that the states of $X$ have "finite energy" so that they decay at
$\infty$ sufficiently fast.

We give a formal definition of solitary wave:

\begin{definition}
\label{solw} A state $\mathbf{u}(x)\in X$ is called solitary wave if%
\[
\left\vert \gamma_{t}\mathbf{u}(x)\right\vert =f(x-vt).
\]
In particular, if $v=0,$ then $\mathbf{u}\left(  x\right)  $ is called
standing wave.
\end{definition}

For example, consider a solution of a field equation having the following
form
\begin{equation}
\mathbf{u}\left(  t,x\right)  =u_{0}(x-\mathbf{v}t)e^{i(\mathbf{v\cdot
x-}\omega t)};\ u_{0}\in L^{2}(\mathbb{R}^{N});
\end{equation}
then, for every $t,$ $\mathbf{u}\left(  t,x\right)  $ is a solitary wave.

The solitons are solitary waves characterized by some form of stability. To
define them at this level of abstractness, we need to recall some well known
notions in the theory of dynamical systems.

A set $\Gamma\subset X$ is called \textit{invariant} if $\forall\mathbf{u}%
\in\Gamma,\forall t\in\mathbb{R},\ \gamma_{t}\mathbf{u}\in\Gamma.$

\begin{definition}
Let $\left(  X,d\right)  $ be a metric space and let $\left(  X,\gamma\right)
$ be a dynamical system. An invariant set $\Gamma\subset X$ is called stable,
if $\forall\varepsilon>0,$ $\exists\delta>0,\;\forall\mathbf{u}\in X$,
\[
d(\mathbf{u},\Gamma)\leq\delta,
\]
implies that
\[
\forall t\in\mathbb{R},\text{ }d(\gamma_{t}\mathbf{u,}\Gamma)\leq\varepsilon.
\]

\end{definition}

\bigskip

Let $G$ be the group induced by the translations in $\mathbb{R}^{N},$ namely,
for every $\tau\in\mathbb{R}^{N},\ $the tranformation $g_{\tau}\in G$ is
defined as follows:
\begin{equation}
\left(  g_{\tau}\mathbf{u}\right)  \left(  x\right)  =\mathbf{u}\left(
x-\tau\right)  .\label{ggg}%
\end{equation}

A subset $\Gamma\subset X$ is called $G$-invariant if
\[
\forall\mathbf{u}\in\Gamma,\ \forall\tau\in\mathbb{R}^{N},\ g_{\tau}%
\mathbf{u}\in\Gamma.
\]

\begin{definition}
A closed $G$-invariant set $\Gamma\subset X$ is called $G$-compact if for any
sequence $\mathbf{u}_{n}(x)$ in $\Gamma$ there is a sequence $\tau_{n}%
\in\mathbb{R}^{N},$ such that $\mathbf{u}_{n}(x-\tau_{n})$ has a converging subsequence.
\end{definition}

Now we are ready to give the definition of soliton:

\begin{definition}
\label{ds} A standing wave $\mathbf{u}(x)$ is called (standing) soliton if
there is an invariant set $\Gamma$ such that

\begin{itemize}
\item (i) $\forall t,\ \gamma_{t}\mathbf{u}(x)\in\Gamma,$

\item (ii) $\Gamma$ is stable,

\item (iii) $\Gamma$ is $G$-compact.
\end{itemize}
\end{definition}

Usually, in the literature, the kind of stability described by the above
definition is called \textit{orbital stability}.

\begin{remark}
The above definition needs some explanation. For simplicity, we assume that
$\Gamma$ is a manifold (actually, it is possible to prove that this is the
generic case if the problem is formulated in a suitable function space). Then
(iii) implies that $\Gamma$ is finite dimensional. Since $\Gamma$ is
invariant, $\mathbf{u}_{0}\in\Gamma\Rightarrow\gamma_{t}\mathbf{u}_{0}%
\in\Gamma$ for every time. Thus, since $\Gamma$ is finite dimensional, the
evolution of $\mathbf{u}_{0}$ is described by a finite number of parameters$.$
Thus the dynamical system $\left(  \Gamma,\gamma\right)  $\ behaves as a point
in a finite dimensional phase space. By the stability of $\Gamma$, a small
perturbation of $\mathbf{u}_{0}$ remains close to $\Gamma.$ However, in this
case, its evolution depends on an infinite number of parameters. Thus, this
system appears as a finite dimensional system with a small perturbation. Since
$\dim(G)=N$, $\dim\left(  \Gamma\right)  \geq N$ and hence, the
\textquotedblright state\textquotedblright\ of a soliton is described by $N$
parameters which define its position and, may be, other parameters which
define its \textquotedblright internal state\textquotedblright.
\end{remark}

\begin{remark}
We recal that (NKGM) are defined by a Lagrangian which is invariant under the
action of the Lorentz group. If $\mathbf{u}_{0}$ is a stationary wave, it is
possible to obtain a travelling wave just making a Lorentz boost (see e.g.
\cite{befogranas} or \cite{milano}). More precisely, let $T_{\mathbf{v}} $ be
the representation of a Lorentz boost relative to our system and let
\[
\mathbf{u}(t,x)=\gamma_{t}\mathbf{u}_{0}(x)
\]
be the evolution of our standing wave $\mathbf{u}_{0}(x);$ then%
\[
\mathbf{u}^{\prime}(t^{\prime},x^{\prime}):=T_{\mathbf{v}}\mathbf{u}(t,x)
\]
is a solution of our equation which moves in time with velocity $\mathbf{v}.$
In \cite{milano} you can see the details and how this principle works in some
particular cases. Obviously, if $\mathbf{u}_{0}$ is a standing soliton,
$\mathbf{u}_{\mathbf{v}}$ is orbitally stable and hence it is a travelling soliton.
\end{remark}

\begin{remark}
\label{qua}The stability of $\Gamma\subset X$ depends on the metric of $X;$
however, if $X$ is a finite dimensional vector space, all the metric (induced
by a norm) are topologically equivalent; hence in this case the stability of
$\Gamma\subset X$ is independent of the metric. This is not the case when $X$
is infinite dimensional. In this case the choice of the "right" metric is a
delicate problem which depends on mathematical and physical considerations. In
many cases, we have that
\begin{equation}
\left\{  \text{energy}\right\}  =\left\{  \text{positive quadratic
form}\right\}  +\left\{  \text{higher order terms}\right\}  .\label{nice}%
\end{equation}
In this case, usually, it is a good choice to use the "norm of the energy":%
\[
\left\Vert \cdot\right\Vert :=\sqrt{\left\{  \text{positive quadratic
form}\right\}  }%
\]

\end{remark}

\subsection{Integrals of motion and hylomorphic solitons\label{im}}

The existence and the properties of hylomorphic solitons are guaranteed by the
interplay between \textit{energy }$E$ and another integral of motion which, in
the general case, is called \textit{hylenic charge} and it will be denoted
by\textit{\ }$C.$ Notice that the first integrals can be considered as
functionals defined on the phase space $X.$

In this section, we shall prove some abstract theorems which guarantee the
existence of hylomorphic solitons. In this paper we will be interested in the
case in which $G$ is the group of translations defined by (\ref{ggg}); however
the abstract theorems hold true for any (locally compact) Lie group; similarly
you do not have to assume that $E(\mathbf{u})$ and $C(\mathbf{u})$ are the
energy and the hylenic charge.

Before stating the abstract theorems, we need some definitions:

\begin{definition}
Let $G$ be a group acting on $X.$ A sequence $\mathbf{u}_{n}$ in $X$ is called
$G$\emph{-compact }if we can extract a subsequence $\mathbf{u}_{n_{k}} $ such
that there exists a sequence $g_{k}\in G$ such that $g_{k}\mathbf{u}_{n_{k}}$
is convergent.
\end{definition}

\begin{definition}
A functional $J$ on $X$ is called $G$\emph{-invariant} if
\[
\forall g\in G,\text{ }\forall\mathbf{u}\in X,\ J\left(  g\mathbf{u}\right)
=J\left(  \mathbf{u}\right)
\]

\end{definition}

\begin{definition}
A functional $J$ on $X$ is called $G$\emph{-compact} if any minimizing
sequence $\mathbf{u}_{n}$ is\emph{\ }$G$-compact.
\end{definition}

\begin{remark}
Clearly, a $G$-compact functional admits a minimizer. Moreover, if $J$ is
$G$-invariant and $\mathbf{u}_{0}$ is a minimizers, then $\left\{
g\mathbf{u_{0}\ |\ }g\in G\right\}  $ is a set of minimizer; so, if $G$ is not
compact (as in the case of the translations group) and if its action is free,
then the set of minimizer is not compact. This fact adds an extra difficulty
to this kind of problems.
\end{remark}

We make the following (abstract) assumptions on the dynamical system
$(X,\gamma)$:

\begin{itemize}
\item (EC-1) there are two prime integrals $E(\mathbf{u})$ and $C(\mathbf{u}%
).$

\item (EC-2) $E(\mathbf{u})$ and $C(\mathbf{u})$ are $G$-invariant.
\end{itemize}

\begin{theorem}
\label{astraco}Assume (EC-1) and (EC-2). For $\mathbf{u}\in X$ and
$e_{0},c_{0}\in$ $\mathbb{R}$, we set%
\begin{equation}
V\left(  \mathbf{u}\right)  =\left(  E\left(  \mathbf{u}\right)
-e_{0}\right)  ^{2}+\left(  C\left(  \mathbf{u}\right)  -c_{0}\right)  ^{2}.
\end{equation}
If $V$ is$\ G$-compact and
\begin{equation}
\Gamma=\left\{  \mathbf{u}\in X:E(\mathbf{u})=e_{0},\ C(\mathbf{u}%
)=c_{0}\right\}  \neq\varnothing,\label{gamma}%
\end{equation}
then every $\mathbf{u}\in\Gamma$ is a soliton.
\end{theorem}

\begin{definition}
A soliton which fulfills the assumptions of Th. \ref{astraco} is called
\emph{hylomorphic.}
\end{definition}

In order to prove Th. \ref{astraco} we need the (well known) Liapunov theorem
in following form:

\begin{theorem}
\label{propV}Let $\Gamma$ be an invariant set and assume that there exists a
differentiable function $V$ (called a Liapunov function) such that

\begin{itemize}
\item (a) $V(\mathbf{u})\geq0$ and\ $V(\mathbf{u})=0\Leftrightarrow u\in
\Gamma$

\item (b) $\partial_{t}V(\gamma_{t}\left(  \mathbf{u}\right)  )\leq0$

\item (c) $V(\mathbf{u}_{n})\rightarrow0\Leftrightarrow d(\mathbf{u}%
_{n},\Gamma)\rightarrow0.$

\noindent Then $\Gamma$ is stable.
\end{itemize}
\end{theorem}

\textbf{Proof. }For completeness, we give a proof of this well known result.
Arguing by contradiction, assume that $\Gamma,$ satisfying the assumptions of
Th. \ref{propV}, is not stable. Then there exists $\varepsilon>0$ and
sequences $\mathbf{u}_{n}\in X$ and $t_{n}>0$ such that
\begin{equation}
d(\mathbf{u}_{n},\Gamma)\rightarrow0\text{ and }d(\gamma_{t_{n}}\left(
\mathbf{u}_{n}\right)  ,\Gamma)>\varepsilon.\label{bingo}%
\end{equation}
Then we have%
\[
d(\mathbf{u}_{n},\Gamma)\rightarrow0\Longrightarrow V(\mathbf{u}%
_{n})\rightarrow0\Longrightarrow V(\gamma_{t_{n}}\left(  \mathbf{u}%
_{n}\right)  )\rightarrow0\Longrightarrow d(\gamma_{t_{n}}\left(
\mathbf{u}_{n}\right)  ,\Gamma)\rightarrow0
\]
where the first and the third implications are consequence of property (c).
The second implication follows from property (b). Clearly, this fact
contradicts (\ref{bingo}).

$\square$

\bigskip

Now we are ready to prove Theorem \textbf{\ref{astraco}}

\bigskip

\textbf{Proof of Th. \ref{astraco}}: We have to prove that $\Gamma$ in
(\ref{gamma}) satisfies (i),(ii) and (iii) of Def. (\ref{ds}). The property
(iii), namely the fact that $\Gamma$ is G-compact, is a trivial consequence of
the fact that $\Gamma$ is the set of minimizers of a G-compact functional $V.$
The invariance property (i) is clearly satisfied since $E$ and $C$ are
constants of the motion. It remains to prove (ii), namely that $\Gamma$ is
stable. To this end we shall use Th. \ref{propV}. So we need to show that
$V(\mathbf{u})$ satisfies (a), (b) and (c). Statements (a) and (b) are
trivial. Now we prove (c). First we show the implication $\Rightarrow.$ Let
$\mathbf{u}_{n}$ be a sequence such that $V(\mathbf{u}_{n})\rightarrow0. $ By
contradiction we assume that $d(\mathbf{u}_{n},\Gamma)\nrightarrow0,$ namely
that there is a subsequence $\mathbf{u}_{n}^{^{\prime}}$ such that
\begin{equation}
d(\mathbf{u}_{n}^{\prime},\Gamma)\geq a>0.\label{kaka}%
\end{equation}
Since $V(\mathbf{u}_{n})\rightarrow0$ also $V(\mathbf{u}_{n}^{\prime
})\rightarrow0,$ and, since $V$ is $G$ compact, there exists a sequence
$g_{n}$ in $G$ such that, for a subsequence $\mathbf{u}_{n}^{\prime\prime}$,
we have $g_{n}\mathbf{u}_{n}^{\prime\prime}\rightarrow\mathbf{u}_{0}.$ Then
\[
d(\mathbf{u}_{n}^{\prime\prime},\Gamma)=d(g_{n}\mathbf{u}_{n}^{\prime\prime
},\Gamma)\leq d(g_{n}\mathbf{u}_{n}^{\prime\prime},\mathbf{u}_{0})\rightarrow0
\]
and this contradicts (\ref{kaka}).

Now we prove the other implication $\Leftarrow.$ Let $\mathbf{u}_{n}$ be a
sequence such that $d(\mathbf{u}_{n},\Gamma)\rightarrow0,$ then there exists
$\mathbf{v}_{n}\in\Gamma$ s.t.
\begin{equation}
d(\mathbf{u}_{n},\Gamma)\geq d(\mathbf{u}_{n},\mathbf{v}_{n})-\frac{1}%
{n}.\label{triplo}%
\end{equation}

Since $V$ is G-compact, also $\Gamma$ is G-compact; so, for a suitable
sequence $g_{n}$, we have $g_{n}\mathbf{v}_{n}\rightarrow\mathbf{\bar{w}}%
\in\Gamma.$ We get the conclusion if we show that $V(\mathbf{u}_{n}%
)\rightarrow0.$ We have by (\ref{triplo}), that $d(\mathbf{u}_{n}%
,\mathbf{v}_{n})\rightarrow0$ and hence $d(g_{n}\mathbf{u}_{n},g_{n}%
\mathbf{v}_{n})\rightarrow0$ and so, since $g_{n}\mathbf{v}_{n}\rightarrow
\mathbf{\bar{w},}$ we have $g_{n}\mathbf{u}_{n}\rightarrow\mathbf{\bar{w}}%
\in\Gamma.$ Therefore, by the continuity of $V$ and since $\mathbf{\bar{w}}%
\in\Gamma,$ we have $V\left(  g_{n}\mathbf{u}_{n}\right)  \rightarrow V\left(
\mathbf{\bar{w}}\right)  =0$ and we can conclude that $V\left(  \mathbf{u}%
_{n}\right)  \rightarrow0.$

$\square$

\bigskip

In the cases in which we are interested, $X$ is an infinite dimensional
manifold; then if you choose generic $e_{0}$ and $c_{0},$ $V$ is not
$G$-compact since the set $\Gamma=\left\{  \mathbf{u}\in X:E(\mathbf{u}%
)=e_{0},\ C(\mathbf{u})=c_{0}\right\}  $ has codimension 2.

The following theorem will be useful to determine $e_{0}$ and $c_{0}$ in such
a way that $V$ is $G$-compact and hence to prove the existence of solitons by
using Theorem \ref{astraco}. It is based on a penalization of the so called
\textit{hylomorphy ratio} $\frac{E(\mathbf{u})}{C(\mathbf{u})}. $

\begin{theorem}
\label{astra} Assume that the dynamical system $(X,\gamma)$ satisfies (EC-1)
and (EC-2). Moreover we set%
\[
J(\mathbf{u})=\frac{E(\mathbf{u})}{\left\vert C(\mathbf{u})\right\vert
}+\delta E(\mathbf{u})^{2}%
\]
where $\delta$ is a positive constant and assume that $J$ is $G$-compact. Then
$J(\mathbf{u})$ has a minimizer $\mathbf{u}_{0}.$ Moreover, if we set
$e_{0}=E(\mathbf{u}_{0}),$ $c_{0}=C(\mathbf{u}_{0}),$ any $\mathbf{u}\in X$
such that $E(\mathbf{u})=e_{0},$ $C(\mathbf{u})=c_{0},$ is an (hylomorphic)
soliton for $(X,\gamma).$
\end{theorem}

\textbf{Proof. }Let $\mathbf{u}_{n}$ be a minimizing sequence of $J.$ $J$ is
$G$-compact, then,for a suitable subsequence $\mathbf{u}_{n_{k}}$ and a
suitable sequence $g_{k}$, we get $g_{k}\mathbf{u}_{n_{k}}\rightarrow
\mathbf{u}_{0}.$ Clearly $\mathbf{u}_{0}$ is a minimizer of $J.$ Now set
$e_{0}=E(\mathbf{u}_{0}),$ $c_{0}=\left\vert C(\mathbf{u}_{0})\right\vert $
and
\begin{equation}
V\left(  \mathbf{u}\right)  =\left(  E\left(  \mathbf{u}\right)
-e_{0}\right)  ^{2}+\left(  C\left(  \mathbf{u}\right)  -c_{0}\right)
^{2}.\label{liap}%
\end{equation}
We show that $V$ is $G$-compact: let $\mathbf{w}_{n}$ be a minimizing sequence
for $V,$ then $V\left(  \mathbf{w}_{n}\right)  \rightarrow0$ and consequently
$E\left(  \mathbf{w}_{n}\right)  \rightarrow e_{0}$ and $C\left(
\mathbf{w}_{n}\right)  \rightarrow c_{0}$. Now, since
\[
\min J=J(\mathbf{u}_{0})=\frac{e_{0}}{c_{0}}+\delta e_{0}^{2},
\]
we have that $\mathbf{w}_{n}$ is a minimizing minimizing sequence also for
$J.$ Then, since $J$ is $G$-compact, we get
\begin{equation}
\mathbf{w}_{n}\ \text{is}\ G\text{-compact}.\label{paracula}%
\end{equation}
So we conclude that $V$ is $G$-compact and hence the conclusion follows by
using Theorem \ref{astraco}.

$\square$

\bigskip

\begin{remark}
The reason why $J(\mathbf{u})$ has such an awkward form depends on the fact
that in our concrete applications is just this form which guarantees the $G$-compactness.
\end{remark}

\section{The Nonlinear Klein-Gordon-Maxwell equations (NKGM)}

\subsection{The Klein-Gordon-Maxwell equations as Abelian gauge theory}

The nonlinear Klein-Gordon equation\ for a complex valued field $\psi,$
defined on the spacetime $\mathbb{R}^{4},$ can be written as follows:%
\begin{equation}
\square\psi+W^{\prime}(\psi)=0\label{KG}%
\end{equation}
where
\[
\square\psi=\frac{\partial^{2}\psi}{\partial t^{2}}-\Delta\psi,\;\;\text{\ }%
\Delta\psi=\frac{\partial^{2}\psi}{\partial x_{1}^{2}}+\frac{\partial^{2}\psi
}{\partial x_{2}^{2}}+\frac{\partial^{2}\psi}{\partial x_{3}^{2}}%
\]
and, with some abuse of notation,
\[
W^{\prime}(\psi)=F^{\prime}(\left\vert \psi\right\vert )\frac{\psi}{\left\vert
\psi\right\vert }%
\]
for some smooth function $F:\left[  0,\infty\right)  \rightarrow\mathbb{R}.$
Hereafter $x=(x_{1},x_{2},x_{3})$ and $t$ will denote the space and time
variables. The field $\psi:$ $\mathbb{R}^{4}\rightarrow\mathbb{C}$ will be
called \textit{matter field}. If $W^{\prime}(s)$ is linear, $W^{\prime
}(s)=m^{2}s,$ $m\neq0,$ equation (\ref{KG}) reduces to the Klein-Gordon
equation. We assume that
\begin{equation}
W(s)=\frac{m^{2}}{2}s^{2}+N(s),\ \text{ }m>0,\ N(s)=o(s^{2}).\label{NN}%
\end{equation}

Consider the Abelian gauge theory in $\mathbb{R}^{4}$ equipped with the
Minkowski metric and described by the Lagrangian density (see
e.g.\cite{befogranas}, \cite{yangL})
\begin{equation}
\mathcal{L}=\mathcal{L}_{0}+\mathcal{L}_{1}-W(\left\vert \psi\right\vert
)\label{marisa}%
\end{equation}
where
\[
\mathcal{L}_{0}=\frac{1}{2}\left(  \left\vert D_{\varphi}\psi\right\vert
^{2}-\left\vert D_{\mathbf{A}}\psi\right\vert ^{2}\right)
\]%
\[
\mathcal{L}_{1}=\frac{1}{2}\left(  \left\vert \partial_{t}\mathbf{%
A%
}+\nabla\phi\right\vert ^{2}-\left\vert \nabla\times\mathbf{A}\right\vert
^{2}\right)  .
\]
Here $q$ denotes a positive parameter which, in some physical model,
represents the unit electric charge, $\nabla\times$ and $\nabla$ denote
respectively the curl and the gradient operators;
\[
\mathbf{%
A%
}=\mathbf{%
\mathbf{(}%
}A_{1},A_{2},A_{3}\mathbf{%
)%
}\in\mathbb{R}^{3}\text{ and }\phi\in\mathbb{R}%
\]
are the gauge potentials;
\[
D_{\varphi}\psi=\left(  \partial_{t}+iq\phi\right)  \psi
\]
is the covariant derivatives with respect to the $t$ variable, and
\[
D_{\mathbf{A}}\psi=\left(  \nabla-iq\mathbf{A}\right)  \psi
\]
is the covariant derivatives with respect to the $x$ variable (see for example
\cite{befogranas} and \cite{yangL}).

Now consider the total action
\begin{equation}
\mathcal{S}=\int\left(  \mathcal{L}_{0}+\mathcal{L}_{1}-W(\left\vert
\psi\right\vert )\right)  dxdt.
\end{equation}

Making the variation of $\mathcal{S}$ with respect to $\psi,$ $\phi$ and
$\mathbf{A,}$ we get the system of the so called
Nonlinear-Klein-Gordon-Maxwell equations (NKGM):
\begin{equation}
D_{\varphi}^{2}\psi-D_{\mathbf{A}}^{2}\psi+W^{\prime}(\psi)=0\label{e1}%
\end{equation}%
\begin{equation}
\nabla\cdot\left(  \partial_{t}\mathbf{%
A%
}+\nabla\phi\right)  =q\operatorname{Re}\left(  iD_{\varphi}\psi\overline
{\psi}\right) \label{e2}%
\end{equation}%
\begin{equation}
\nabla\times\left(  \nabla\times\mathbf{A}\right)  +\partial_{t}\left(
\partial_{t}\mathbf{%
A%
}+\nabla\phi\right)  =q\operatorname{Re}\left(  iD_{\mathbf{A}}\psi
\overline{\psi}\right)  .\label{e3}%
\end{equation}
Here $\nabla\cdot$ denotes the divergence operator.

\subsection{General features of the Klein-Gordon-Maxwell equations}

If we make the following change of variables:
\begin{equation}
\mathbf{E=-}\left(  \frac{\partial\mathbf{%
A%
}}{\partial t}+\nabla\phi\right) \label{pos1}%
\end{equation}%
\begin{equation}
\mathbf{H}=\nabla\times\mathbf{A}\label{pos2}%
\end{equation}%
\begin{equation}
\rho=-q\operatorname{Re}\left(  iD_{\varphi}\psi\overline{\psi}\right)
\label{caricona}%
\end{equation}%
\begin{equation}
\mathbf{j}=q\operatorname{Re}\left(  iD_{\mathbf{A}}\psi\overline{\psi
}\right)  ,\label{tar}%
\end{equation}
we see that (\ref{e2}) and (\ref{e3}) are the second couple of the Maxwell
equations (Gauss and Ampere laws) with respect to a matter distribution whose
electric charge and current densities are respectively $\rho$ and
$\mathbf{j.}$ These equations can be written as follows:
\begin{equation}
\nabla\cdot\mathbf{E}=\rho\tag{\textsc{gauss}}\label{gauss}%
\end{equation}%
\begin{equation}
\nabla\times\mathbf{H}-\frac{\partial\mathbf{E}}{\partial t}=\mathbf{j}%
.\tag{\textsc{ampere}}\label{ampere}%
\end{equation}
Equations (\ref{pos1}) and (\ref{pos2}) give rise to the first couple of the
Maxwell equations:
\begin{equation}
\nabla\times\mathbf{E}+\frac{\partial\mathbf{H}}{\partial t}%
=0\tag{\textsc{faraday}}\label{faraday}%
\end{equation}%
\begin{equation}
\nabla\cdot\mathbf{H}=0.\tag{\textsc{nomonopole}}\label{monopole}%
\end{equation}

Sometimes it is useful to give a different form to these equations; if we
write $\psi$ in polar form
\begin{equation}
\psi(x,t)=u(x,t)\,e^{iS(x,t)},\;\;u\geq0,\;\;S\in\mathbb{R}/2\pi
\mathbb{Z}\label{giulia}%
\end{equation}
equation (\ref{e1}) can be split in the two following ones
\begin{equation}
\square u+W^{\prime}(u)+\left[  \left\vert \nabla S-q\mathbf{A}\right\vert
^{2}-\left(  \frac{\partial S}{\partial t}+q\phi\right)  ^{2}\right]
\,u=0\label{e1+}%
\end{equation}%
\begin{equation}
\frac{\partial}{\partial t}\left[  \left(  \frac{\partial S}{\partial t}%
+q\phi\right)  u^{2}\right]  -\nabla\cdot\left[  \left(  \nabla S-q\mathbf{A}%
\right)  u^{2}\right]  =0.\label{e1cont}%
\end{equation}
Observe that, using the polar form (\ref{giulia}), (\ref{caricona}) and
(\ref{tar}) become%
\[
\rho=-q\left(  \frac{\partial S}{\partial t}+q\phi\right)  u^{2},\text{
}\mathbf{j}=q\left(  \nabla S-q\mathbf{A}\right)  u^{2}.
\]
Then equations (\ref{e1+}) and (\ref{e1cont}), using the variables
$\mathbf{j}$ and $\rho,$ can be written as follows:
\begin{equation}
\square u+W^{\prime}(u)+\frac{\mathbf{j}^{2}-\rho^{2}}{q^{2}u^{3}%
}=0\tag{\textsc{matter}}\label{materia}%
\end{equation}%
\begin{equation}
\frac{\partial\rho}{\partial t}+\nabla\cdot\mathbf{j}=0.\label{continuit�}%
\end{equation}
Equation (\ref{continuit�}) is the charge continuity equation.

Notice that equation (\ref{continuit�}) is also a consequence of
(\ref{gauss}) and (\ref{ampere}) and hence it can be eliminated. Thus
equations (\ref{e1},\ref{e2},\ref{e3}) are equivalent to equations
(\ref{gauss}, \ref{ampere}, \ref{faraday}, \ref{monopole}, \ref{materia}).

In conclusion, an Abelian gauge theory, via equations (\ref{gauss},
\ref{ampere}, \ref{faraday}, \ref{monopole}, \ref{materia}), provides a model
of interaction of the matter field $\psi$ with the electromagnetic field
$(\mathbf{E},\mathbf{H})$.

Observe that the Lagrangian (\ref{marisa}) is invariant with respect to the
gauge transformations%
\begin{equation}
\psi\rightarrow e^{iq\chi}\psi\label{ga}%
\end{equation}%
\begin{equation}
\phi\rightarrow\phi-\partial_{t}\chi\label{ge}%
\end{equation}%
\begin{equation}
\mathbf{A\rightarrow A}+\nabla\chi\label{gi}%
\end{equation}
where $\chi\in C^{\infty}\left(  \mathbb{R}^{4}\right)  $.

So, our equations are gauge invariant; if we use the variables $u,\rho
,\mathbf{j},\mathbf{E},$ $\mathbf{H,}$ this fact can be checked directly since
these variables are gauge invariant. Actually, equations (\ref{gauss},
\ref{ampere}, \ref{faraday}, \ref{monopole}, \ref{materia}) are the gauge
invariant formulation of equations (\ref{e1},\ref{e2},\ref{e3}).

\subsection{The phase space $X$}

Noether's theorem states that any invariance for a one-parameter group of the
Lagrangian implies the existence of an integral of motion (see e.g.
\cite{Gelfand}, \cite{milano}). In particular the invariants, which are
relevant for us, are the energy and the charge, which, in the gauge invariant
variables, take the following form (for the explicit computation of $E$ see
e.g. \cite{befogranas})%
\[
E\left(  \mathbf{u}\right)  =\frac{1}{2}\int\left(  \left\vert \partial
_{t}u\right\vert ^{2}+\left\vert \nabla u\right\vert ^{2}+\frac{\rho
^{2}+\mathbf{j}^{2}}{q^{2}u^{2}}+\mathbf{E}^{2}+\mathbf{H}^{2}\right)  dx+\int
W(u)dx
\]%
\[
C_{el}=\int\rho\ dx.
\]
Observe that $C_{el}$ is the electric charge and $C=\frac{C_{el}}{q}$ is the
hylenic charge (as defined e.g. in \cite{milano}).

The term $\left(  \rho^{2}+\mathbf{j}^{2}\right)  /u^{2}$ is singular and the
energy does not have the form (\ref{nice}). In order to avoid this problem, it
is convenient to introduce new gauge invariant variables which eliminate this
singularity:%
\[
\theta=\frac{-\rho}{qu};\ \Theta=\frac{\mathbf{j}}{qu}.
\]
Using these new variables the energy takes the form:%
\begin{align*}
E\left(  \mathbf{u}\right)   & =\frac{1}{2}\int\left(  \left\vert \partial
_{t}u\right\vert ^{2}+\left\vert \nabla u\right\vert ^{2}+\theta^{2}%
+\Theta^{2}+\mathbf{E}^{2}+\mathbf{H}^{2}\right)  dx+\int W(u)dx\\
& =\frac{1}{2}\int\left[  \left\vert \partial_{t}u\right\vert ^{2}+\left\vert
\nabla u\right\vert ^{2}+m^{2}u^{2}+\theta^{2}+\Theta^{2}+\mathbf{E}%
^{2}+\mathbf{H}^{2}\right]  +\int N(u).
\end{align*}

Thus, we can construct the space $X$ and its metric taking into account the
suggestions of remark \ref{qua}. The generic point in the phase space is given
by%
\[
\mathbf{u}=\left(  u,\hat{u},\theta,\Theta,\mathbf{E},\mathbf{H}\right)
\]
where $\hat{u}=\partial_{t}u$ is considered as independent variable; the phase
space is given by%
\begin{equation}
X=\left\{  \mathbf{u}\in\mathcal{H}:\nabla\cdot\mathbf{E}=-q\theta
u,\ \nabla\cdot\mathbf{H}=0\right\} \label{xxx}%
\end{equation}
where $\mathcal{H}$ is the Hilbert space of the functions
\[
\mathbf{u}=\left(  u,\hat{u},\theta,\Theta,\mathbf{E},\mathbf{H}\right)  \in
H^{1}\left(  \mathbb{R}^{3}\right)  \times L^{2}\left(  \mathbb{R}^{3}\right)
^{11}%
\]
equipped with the norm defined by the quadratic part of the energy:%
\begin{equation}
\left\Vert \mathbf{u}\right\Vert ^{2}=\int\left[  \hat{u}^{2}+\left\vert
\nabla u\right\vert ^{2}+m^{2}u^{2}+\theta^{2}+\Theta^{2}+\mathbf{E}%
^{2}+\mathbf{H}^{2}\right]  dx\label{nor}%
\end{equation}

In these new variables the energy and the hylenic charge become two continuous
functionals on $X$ having the form%
\begin{equation}
E\left(  \mathbf{u}\right)  =\frac{1}{2}\int\left(  \left\vert \hat{u}%
^{2}\right\vert ^{2}+\left\vert \nabla u\right\vert ^{2}+\theta^{2}+\Theta
^{2}+\mathbf{E}^{2}+\mathbf{H}^{2}\right)  dx+\int W(u)dx\label{placo}%
\end{equation}

\begin{equation}
C\left(  \mathbf{u}\right)  =\int\theta u\ dx.\label{plat}%
\end{equation}

Our equations (\ref{materia}, \ref{gauss}, \ref{ampere}, \ref{faraday},
\ref{monopole}) become%
\begin{align}
\square u+W^{\prime}(u)+\frac{\Theta^{2}-\theta^{2}}{u}  & =0\nonumber\\
\nabla\cdot\mathbf{E}  & =-q\theta u\nonumber\\
\nabla\times\mathbf{H}-\frac{\partial\mathbf{E}}{\partial t}  & =q\Theta
u\label{equazioni}\\
\nabla\times\mathbf{E}+\frac{\partial\mathbf{H}}{\partial t}  & =0\nonumber\\
\nabla\cdot\mathbf{H}  & =0.\nonumber
\end{align}

\begin{remark}
In the following we shall assume that the Cauchy problem for (NKGM) is well
posed in $X.$ Actually, in the literature there are few results relative to
this problem (we know only \cite{kleinerman}) and we do not know which are the
assumptions that $W$ should satisfy. Also, we refer to \cite{befo11} for a
discussion and some partial result on this issue.
\end{remark}

\section{The existence result}

\subsection{Statement of the main results}

We make the following assumptions:

\begin{itemize}
\item (W-i) \textbf{(Positivity}) $W(s)\geq0$

\item (W-ii) \textbf{(Nondegeneracy}) $W=$ $W(s)$ ( $s\geq0)$ is $C^{2}$ near
the origin with $W(0)=W^{\prime}(0)=0;\;W^{\prime\prime}(0)=m^{2}$\ $>0$

\item (W-iii) \textbf{(Hylomorphy}) $\exists\bar{s}>0\ $and $\alpha\in\left(
0,m\right)  $ such that $W(\bar{s})\leq\frac{1}{2}\alpha^{2}\bar{s}^{2}$

\item (W-iiii)\textbf{(Growth condition}) There are constants $a,b>0,$ $6>p>2
$ s.t. $|N^{\prime}(s)|\ \leq as^{p-1}+bs^{2-\frac{2}{p}}$ where $N$ id
defined by eq. (\ref{NN}).
\end{itemize}

Here there are some comments on assumptions (W-i), (W-ii), (W-iii), (W-iiii).

(W-i) Clearly (see (\ref{placo})) (W-i) implies that the energy is positive;
if this condition does not hold, it is possible to have solitary waves, but
not hylomorphic waves (cf. the discussion in section 4.2 of \cite{befo08}).

(W-ii) In order to have solitary waves it is necessary to have $W^{\prime
\prime}(0)\geq0.$ There are some results also when $W^{\prime\prime}(0)=0 $
(null-mass case, see e.g. \cite{Beres-Lions}), however the most interesting
situation occurs when $W^{\prime\prime}(0)>0.$

(W-iii) This is the crucial assumption which characterizes the potentials
which might produce hylomorphic solitons. As we will see, this assumption
permits to have states $\mathbf{u}$ with hylomorphy ratio $\frac
{E(\mathbf{u)}}{C(\mathbf{u)}}<m$. By this assumption there exists $s_{0}$
such that $\;N(s_{0})<0$\ 

(W-iiii) This assumption contains the usual growth condition at infinity which
guarantees the $C^{1}$ regularity of the functional. Moreover it implies that
$\left\vert N^{\prime}(s)\right\vert $ $=O($ $s^{2-\frac{2}{p}})$ for $s$ small.

\bigskip

We have the following results:

\begin{theorem}
\label{main-theorem} Assume that (W-i),(W-ii),(W-iii),(W-iiii) hold, then
there exists $\bar{q}$ such that for every $q\in\left[  0,\bar{q}\right]  $,
equations (\ref{equazioni}) have a continuous family $\mathbf{u}_{\delta}$
($\delta\in\left(  0,\bar{\delta}\left(  q\right)  \right)  $) of independent,
hylomorphic solitons (two solitons $\mathbf{u}_{\delta_{1}},\mathbf{u}%
_{\delta_{2}}$ are called independent if $\mathbf{u}_{\delta_{1}}\neq
g\mathbf{u}_{\delta_{2}}$ for every $g\in G$).

\begin{theorem}
\label{imp}The solitons $\mathbf{u}_{\delta}=\left(  u_{\delta},\hat
{u}_{\delta},\theta_{\delta},\Theta_{\delta},\mathbf{E}_{\delta}%
,\mathbf{H}_{\delta}\right)  $ in Theorem \ref{main-theorem} are stationary
solutions of (\ref{equazioni}), this means that $\hat{u}_{\delta}%
=\Theta_{\delta}=\mathbf{H}_{\delta}=0,$ $\mathbf{E}_{\delta}=-\nabla
\phi_{\delta}$ and $u_{\delta},\theta_{\delta},\phi_{\delta}$ solve the
equations
\begin{align}
-\Delta u_{\delta}+W^{\prime}(u_{\delta})-\frac{\theta_{\delta}^{2}}%
{u_{\delta}}  & =0\label{stat1}\\
-\Delta\phi_{\delta}  & =-q\theta_{\delta}u_{\delta}\label{stat2}%
\end{align}

\end{theorem}
\end{theorem}

The proofs of Theorem \ref{main-theorem} and of Theorem \ref{imp} will be
given in the next section.

\subsection{Proof of the main results}

First of all we introduce the following notation:%
\begin{equation}
Q=\left\{  x=(x_{1},...,x_{N})\in\mathbb{R}^{N}:0\leq x_{i}%
<1,\ i=1,..,N\right\} \label{qu}%
\end{equation}
\[
Q_{j}=j+Q,\ j\in\mathbb{Z}^{3}%
\]%
\[
X\left(  Q_{j}\right)  =\left\{  \left.  \underset{}{\mathbf{u}}\right\vert
_{Q_{j}}:\mathbf{u\in}X\right\}
\]%
\[
\left\Vert \mathbf{u}\right\Vert _{Q_{j}}^{2}=\int_{Q_{j}}\left[  \hat{u}%
^{2}+\left\vert \nabla u\right\vert ^{2}+m^{2}u^{2}+\theta^{2}+\Theta
^{2}+\mathbf{E}^{2}+\mathbf{H}^{2}\right]  dx
\]

\[
E_{Q_{j}}(\mathbf{u})=\frac{1}{2}\int_{Q_{j}}\left[  \left\vert \hat
{u}\right\vert ^{2}+\theta^{2}+\left\vert \nabla u\right\vert ^{2}+\Theta
^{2}+u^{2}+\mathbf{E}^{2}+\mathbf{H}^{2}\right]  dx+\int_{Q_{j}}W(u)dx
\]%
\[
C_{Q_{j}}(\mathbf{u})=\int_{Q_{j}}\theta u\ dx
\]
and%
\[
\Lambda_{0}=\ \underset{\mathbf{u}\in X}{\inf}\ \frac{\frac{1}{2}\left\Vert
\mathbf{u}\right\Vert _{Q}^{2}}{\left\vert C_{Q}\left(  \mathbf{u}\right)
\right\vert },\text{ }\Lambda_{\ast}=\ \underset{\mathbf{u}\in X}{\inf}%
\ \frac{E\left(  \mathbf{u}\right)  }{\left\vert C\left(  \mathbf{u}\right)
\right\vert }=\ \underset{\mathbf{u}\in X}{\inf}\ \frac{\frac{1}{2}\left\Vert
\mathbf{u}\right\Vert ^{2}+\int N(u)dx}{\left\vert C\left(  \mathbf{u}\right)
\right\vert }%
\]

\begin{lemma}
\label{uno}The following inequality holds:%
\[
\Lambda_{0}\geq m.
\]

\end{lemma}

\textbf{Proof:}%
\begin{align*}
\Lambda_{0}  & =\ \underset{\mathbf{u}\in X}{\inf}\ \frac{\frac{1}%
{2}\left\Vert \mathbf{u}\right\Vert _{Q}^{2}}{\left\vert C_{Q}\left(
\mathbf{u}\right)  \right\vert }\\
& \geq\ \underset{\mathbf{u}\in X}{\inf}\frac{\frac{1}{2}\int_{Q}\left[
\hat{u}^{2}+\left\vert \nabla u\right\vert ^{2}+m^{2}u^{2}+\theta^{2}%
+\Theta^{2}+\mathbf{E}^{2}+\mathbf{H}^{2}\right]  dx}{\int_{Q}\left\vert
\theta u\right\vert \ dx}\\
& \geq\ \underset{\mathbf{u}\in X}{\inf}\frac{\frac{1}{2}\int_{Q}\left[
m^{2}u^{2}+\theta^{2}\right]  dx}{\int_{Q}\left\vert \theta u\right\vert
\ dx}=\ \underset{\mathbf{u}\in X}{\inf}\frac{\int_{Q}m\left\vert u\right\vert
~\left\vert \theta\right\vert dx}{\int_{Q}\left\vert \theta u\right\vert
\ dx}\geq m
\end{align*}

$\square$

\bigskip

The next lemma provides a crucial estimate for the existence of solitons:

\begin{lemma}
\label{due}Let $\alpha,\bar{s},m$ be the positive constants appearing in
(W-iii). Then there is a positive constant $c$ such that for any $0<q<\frac
{c}{\bar{s}}\sqrt{\left(  m-\alpha\right)  ^{3}\alpha}$ we have%
\[
\Lambda_{\ast}<m
\]

\end{lemma}

\textbf{Proof:}\ We set
\begin{equation}
u_{R}=\left\{
\begin{array}
[c]{cc}%
\bar{s} & if\;\;|x|<R\\
0 & if\;\;|x|>R+1\\
\frac{|x|}{R}\bar{s}-(\left\vert x\right\vert -R)\frac{R+1}{R}\bar{s} &
if\;\;R<|x|<R+1
\end{array}
\right. \label{inff}%
\end{equation}
where $R>1.$ Moreover we denote by $\varphi_{R}\in\mathcal{D}^{1,2}$ the
solution of the following equation%
\begin{equation}
\Delta\varphi=-q\alpha u_{R}^{2}.\label{mar}%
\end{equation}

We have%
\begin{align*}
\Lambda_{\ast}  & =\ \underset{\mathbf{u}\in X}{\inf}\ \frac{\frac{1}%
{2}\left\Vert \mathbf{u}\right\Vert ^{2}+\int N(u)dx}{\left\vert C\left(
\mathbf{u}\right)  \right\vert }\\
& =\ \underset{\mathbf{u}\in X}{\inf}\frac{\frac{1}{2}\int\left[  \hat{u}%
^{2}+\left\vert \nabla u\right\vert ^{2}+\theta^{2}+\Theta^{2}+\mathbf{E}%
^{2}+\mathbf{H}^{2}\right]  dx+\int W(u)dx}{\left\vert \int\theta
u\ dx\right\vert }.
\end{align*}

Now remember that $\mathbf{u}=\left(  u,\hat{u},\theta,\Theta,\mathbf{E}%
,\mathbf{H}\right)  $ and take $\mathbf{u}=\mathbf{u}_{R}$ with
\[
\mathbf{u}_{R}=\left(  u_{R},0,\alpha u_{R},0,\nabla\varphi_{R},\mathbf{0}%
\right)  .
\]
By (\ref{xxx}) and (\ref{mar}), $\mathbf{u}_{R}\in X;$ then we have%

\begin{align}
\Lambda_{\ast}  & =\ \underset{\mathbf{u}\in X}{\inf}\ \frac{\frac{1}%
{2}\left\Vert \mathbf{u}\right\Vert ^{2}+\int N(u)dx}{\left\vert C\left(
\mathbf{u}\right)  \right\vert }\leq\frac{\frac{1}{2}\left\Vert \mathbf{u}%
_{R}\right\Vert ^{2}+\int N(u_{R})dx}{C\left(  \mathbf{u}_{R}\right)
}\nonumber\\
& =\frac{\frac{1}{2}\int\left[  \left\vert \nabla u_{R}\right\vert ^{2}%
+\alpha^{2}u_{R}^{2}+\left\vert \nabla\varphi_{R}\right\vert ^{2}\right]
dx+\int W(u_{R})dx}{\alpha\int u_{R}^{2}\ dx}\nonumber\\
& \leq\frac{\frac{1}{2}\int_{\left\vert x\right\vert <R}\left[  \left\vert
\nabla u_{R}\right\vert ^{2}+\alpha^{2}u_{R}^{2}\right]  +\int_{\left\vert
x\right\vert <R}W(u_{R})}{\alpha\int_{\left\vert x\right\vert <R}u_{R}%
^{2}\ dx}\nonumber\\
& +\frac{\frac{1}{2}\int_{R<\left\vert x\right\vert <R+1}\left[  \left\vert
\nabla u_{R}\right\vert ^{2}+\alpha^{2}u_{R}^{2}\right]  +\int_{R<\left\vert
x\right\vert <R+1}W(u_{R})}{\alpha\int_{\left\vert x\right\vert <R}u_{R}%
^{2}\ }+\frac{\frac{1}{2}\int\left\vert \nabla\varphi_{R}\right\vert ^{2}%
}{\alpha\int_{\left\vert x\right\vert <R}u_{R}^{2}\ }\nonumber\\
& =\frac{\frac{1}{2}\int_{\left\vert x\right\vert <R}\alpha^{2}\bar{s}%
^{2}+\int_{\left\vert x\right\vert <R}W(\bar{s})}{\alpha\int_{\left\vert
x\right\vert <R}\bar{s}^{2}}+\frac{c_{1}R^{2}}{\alpha\int_{\left\vert
x\right\vert <R}\bar{s}^{2}}+\frac{\frac{1}{2}\int\left\vert \nabla\varphi
_{R}\right\vert ^{2}}{\alpha\int_{\left\vert x\right\vert <R}\bar{s}^{2}%
\ \ }\nonumber\\
& \leq\alpha+\frac{c_{2}}{\alpha R}+\frac{\frac{1}{2}\int\left\vert
\nabla\varphi_{R}\right\vert ^{2}}{\frac{4}{3}\pi\alpha\bar{s}^{2}R^{3}%
\ }\label{star}%
\end{align}
where the last inequality is a consequence of (W-iii).

In order to estimate the term containing $\varphi_{R}$ in (\ref{star}), we
remember that $\varphi_{R}$ is the solution of (\ref{mar}). Observe that
$u_{R}^{2}$ has radial symmetry and that the electric field outside any
spherically symmetric charge distribution is the same as if all of the charge
were concentrated into a point. So $\left\vert \nabla\varphi_{R}\left(
r\right)  \right\vert $ corresponds to the strength of an electrostatic field
at distance $r,$ created by an electric charge given by
\[
\left\vert C_{el}\right\vert =%
{\displaystyle\int\limits_{\left\vert x\right\vert \leq r}}
q\alpha u_{R}^{2}dx=4\pi%
{\displaystyle\int\limits_{0}^{r}}
q\alpha u_{R}^{2}v^{2}dv
\]
and located at the origin. So we have%
\[
\left\vert \nabla\varphi_{R}\left(  r\right)  \right\vert =\frac{\left\vert
C_{el}\right\vert }{r^{2}}\left\{
\begin{array}
[c]{cc}%
=\frac{4}{3}\pi q\alpha\bar{s}^{2}r & if\ r<R\\
\leq\frac{4}{3}\pi q\alpha\bar{s}^{2}\frac{(R+1)^{3}}{r^{2}} & if\ r\geq R
\end{array}
\right.
\]

Then%
\begin{align*}
\int\left\vert \nabla\varphi_{R}\right\vert ^{2}dx  & \leq c_{3}q^{2}%
\alpha^{2}\bar{s}^{4}\left(  \int_{r<R}r^{4}dr+\int_{r>R}\frac{(R+1)^{6}%
}{r^{2}}dr\right) \\
& \leq c_{4}q^{2}\alpha^{2}\bar{s}^{4}\left(  R^{5}+\frac{(R+1)^{6}}%
{R}\right)  \leq c_{5}q^{2}\alpha^{2}\bar{s}^{4}R^{5}.
\end{align*}
Then%
\begin{equation}
\frac{\frac{1}{2}\int\left\vert \nabla\varphi_{R}\right\vert ^{2}}{\frac{4}%
{3}\pi\alpha\bar{s}^{2}R^{3}\ }\leq c_{6}q^{2}\alpha\bar{s}^{2}R^{2}%
.\label{pa}%
\end{equation}
By (\ref{star}) and (\ref{pa}), we get%

\begin{equation}
\Lambda_{\ast}\leq\alpha+\frac{c_{1}}{\alpha R}+c_{6}q^{2}\alpha\bar{s}%
^{2}R^{2}.\label{quasi}%
\end{equation}

Now set
\[
m-\alpha=2\varepsilon
\]
and take
\[
R=\frac{c_{1}}{\alpha\varepsilon},\text{ }0<q<\sqrt{\frac{\varepsilon
^{3}\alpha}{\bar{s}^{2}c_{1}^{2}c_{6}}}.
\]
With these choices of $R$ and $q,$ a direct calculation shows that%
\begin{equation}
\alpha+\frac{c_{1}}{\alpha R}+c_{6}q^{2}\alpha\bar{s}^{2}R^{2}<m.\label{cucu}%
\end{equation}
Then, by (\ref{quasi}) and (\ref{cucu}), we get that there exists a positive
constant $c$ such that, for $0<q<\frac{c}{\bar{s}}\sqrt{\left(  m-\alpha
\right)  ^{3}\alpha}$, we have
\begin{equation}
\Lambda_{\ast}<m.\label{dopo}%
\end{equation}

$\square$

\bigskip

\begin{lemma}
\label{tre}Consider any sequence
\[
\mathbf{u}_{n}=\mathbf{u}+\mathbf{w}_{n}\in X
\]
where $\mathbf{w}_{n}$ converges weakly to $0.$ Then%
\begin{equation}
E(\mathbf{u}_{n})=E(\mathbf{u})+E(\mathbf{w}_{n})+o(1)\label{mi}%
\end{equation}
and%
\begin{equation}
C(\mathbf{u}_{n})=C(\mathbf{u})+C(\mathbf{w}_{n})+o(1).\label{mu}%
\end{equation}

\end{lemma}

\textbf{Proof. }First of all we introduce the following notation:

As usual $u,w_{n}$ will denote the first components respectively of
$\mathbf{u,w}_{n}\in H^{1}\left(  \mathbb{R}^{3}\right)  \times L^{2}\left(
\mathbb{R}^{3}\right)  ^{11}$.

If $v\in H^{1}\left(  \mathbb{R}^{3}\right)  ,$ we set%
\[
K(v\mathbf{)=}\int N(v)dx
\]

and for any measurable $A\subset\mathbb{R}^{N}$,%
\[
K_{A}(v\mathbf{)=}\int_{A}N(v)dx
\]

We have to show that $\underset{n\rightarrow\infty}{\lim}\left\vert E\left(
\mathbf{u}+\mathbf{w}_{n}\right)  -E\left(  \mathbf{u}\right)  -E\left(
\mathbf{w}_{n}\right)  \right\vert =0.$ By (\ref{nor}), (\ref{placo}) we have
that%
\begin{align*}
& \underset{n\rightarrow\infty}{\lim}\left\vert E\left(  \mathbf{u}%
+\mathbf{w}_{n}\right)  -E\left(  \mathbf{u}\right)  -E\left(  \mathbf{w}%
_{n}\right)  \right\vert \\
& \leq\ \underset{n\rightarrow\infty}{\lim\frac{1}{2}}\left\vert \left\Vert
\mathbf{u}+\mathbf{w}_{n}\right\Vert ^{2}-\left\Vert \mathbf{u}\right\Vert
^{2}-\left\Vert \mathbf{w}_{n}\right\Vert ^{2}\right\vert +\ \underset
{n\rightarrow\infty}{\lim}\left\vert K\left(  u+w_{n}\right)  -K\left(
u\right)  -K\left(  w_{n}\right)  \right\vert .
\end{align*}

Let us consider each piece independently. If $(\cdot,\cdot)$ denotes the inner
product induced by the norm $\left\Vert {}\right\Vert $ we have:%
\[
\underset{n\rightarrow\infty}{\lim}\left\vert \left\Vert \mathbf{u}%
+\mathbf{w}_{n}\right\Vert ^{2}-\left\Vert \mathbf{u}\right\Vert
^{2}-\left\Vert \mathbf{w}_{n}\right\Vert ^{2}\right\vert =\ \underset
{n\rightarrow\infty}{\lim}\left\vert 2\left(  \mathbf{u},\mathbf{w}%
_{n}\right)  \right\vert =0.
\]

Choose $\varepsilon>0$ and $R=R(\varepsilon)>0$ such that
\begin{equation}
\left\vert K_{B_{R}^{c}}\left(  u\right)  \right\vert <\varepsilon\label{bis}%
\end{equation}

$\ $where $\ $%
\[
B_{R}^{c}=\mathbb{R}^{N}-B_{R}\text{ and }B_{R}=\left\{  x\in\mathbb{R}%
^{N}:\left\vert x\right\vert <R\right\}  .
\]
Since $w_{n}\rightharpoonup0$ weakly in $H^{1}\left(  \mathbb{R}^{3}\right)
$, by usual compactness arguments, we have that
\begin{equation}
K_{B_{R}}\left(  w_{n}\right)  \rightarrow0\text{ and }K_{B_{R}}\left(
u+w_{n}\right)  \rightarrow K_{B_{R}}\left(  u\right)  .\label{bibis}%
\end{equation}
Then, by (\ref{bis}) and (\ref{bibis}), we have%

\begin{align}
& \underset{n\rightarrow\infty}{\lim}\left\vert K\left(  u+w_{n}\right)
-K\left(  u\right)  -K\left(  w_{n}\right)  \right\vert \nonumber\\
& =\ \underset{n\rightarrow\infty}{\lim}\left\vert K_{B_{R}^{c}}\left(
u+w_{n}\right)  +K_{B_{R}}\left(  u+w_{n}\right)  -K_{B_{R}^{c}}\left(
u\right)  -K_{B_{R}}\left(  u\right)  -K_{B_{R}^{c}}\left(  w_{n}\right)
-K_{B_{R}}\left(  w_{n}\right)  \right\vert \nonumber\\
& \mathbb{=}\ \underset{n\rightarrow\infty}{\lim}\left\vert K_{B_{R}^{c}%
}\left(  u+w_{n}\right)  -K_{B_{R}^{c}}\left(  u\right)  -K_{B_{R}^{c}}\left(
w_{n}\right)  \right\vert \nonumber\\
& \leq\ \underset{n\rightarrow\infty}{\lim}\left\vert K_{B_{R}^{c}}\left(
u+w_{n}\right)  -K_{B_{R}^{c}}\left(  w_{n}\right)  \right\vert +\varepsilon
.\label{pepe}%
\end{align}

Now observe that, for $\Omega\subset\mathbb{R}^{3},$ we have that
\begin{equation}
K^{\prime}\text{ is bounded from }H^{1}\left(  \Omega\right)  \text{ into its
dual }\left(  H^{1}\left(  \Omega\right)  \right)  ^{\prime}\label{bond}%
\end{equation}
In fact let $v_{n}$ be a bounded sequence in $H^{1}\left(  \Omega\right)  ,$
then, by the growth assumption (W-iiii) we have
\[
|N^{\prime}(v_{n})|\ \leq a\left\vert v_{n}\right\vert ^{p-1}+b\left\vert
v_{n}\right\vert ^{2-\frac{2}{p}}.
\]
Elevating both members to $p^{\prime}=\frac{p}{p-1}$ we get%
\begin{equation}
|N^{\prime}(v_{n})|^{p^{\prime}}\ \leq c_{1}\left\vert v_{n}\right\vert
^{p}+c_{2}\left\vert v_{n}\right\vert ^{2}.\label{bubu}%
\end{equation}

Then, since $v_{n}$ is bounded in $H^{1}\left(  \Omega\right)  $ and by
(\ref{bubu}), we have that $N^{\prime}(v_{n})$ is bounded in $L^{p^{\prime}}$
and then it is bounded also in $\left(  H^{1}\left(  \Omega\right)  \right)
^{\prime}.$ Then (\ref{bond}) is proved.

By the intermediate value theorem and by (\ref{bond}) it is easy to deduce
that there exist $R$ and $M>0$ sufficiently large and $\zeta_{n}\in(0,1)$ such
that
\begin{equation}
\left\vert K_{B_{R}^{c}}\left(  u+w_{n}\right)  -K_{B_{R}^{c}}\left(
w_{n}\right)  \right\vert \leq\left\Vert K_{B_{R}^{c}}^{\prime}\left(
\zeta_{n}u+\left(  1-\zeta_{n}\right)  w_{n}\right)  \right\Vert _{\left(
H^{1}\left(  B_{R}^{c}\right)  \right)  ^{\prime}}\cdot\left\Vert u\right\Vert
_{H^{1}\left(  B_{R}^{c}\right)  }\leq M\cdot\varepsilon\label{fere}%
\end{equation}

Then, by (\ref{pepe}) and (\ref{fere}), we get%
\[
\underset{n\rightarrow\infty}{\lim}\left\vert K\left(  u+w_{n}\right)
-K\left(  u\right)  -K\left(  w_{n}\right)  \right\vert \leq\varepsilon
+M\cdot\varepsilon
\]
Since $\varepsilon$ is arbitrary, this limit is 0. Then we have proved
(\ref{mi}) $.$ The proof of (\ref{mu}) is immediate.

$\square$

\bigskip

Now we choose
\[
0<q<\frac{c}{\bar{s}}\sqrt{\left(  m-\alpha\right)  ^{3}\alpha}.
\]
Then, by lemma \ref{uno} and (\ref{dopo}), we have that
\[
\Lambda_{\ast}<\Lambda_{0}.
\]
So there exists $\mathbf{u}_{0}\in X$ and $b>0$ such that
\[
\frac{E(\mathbf{u}_{0})}{\left\vert C(\mathbf{u}_{0})\right\vert }\leq
\Lambda_{0}-b.
\]
Then we can choose $\delta>0$ such that%
\begin{equation}
\frac{E(\mathbf{u}_{0})}{\left\vert C(\mathbf{u}_{0})\right\vert }+\delta
E(\mathbf{u}_{0})^{2}\leq\Lambda_{0}-\frac{b}{2}\label{kkk}%
\end{equation}
and we define
\begin{equation}
J(\mathbf{u})=\frac{E(\mathbf{u})}{\left\vert C(\mathbf{u})\right\vert
}+\delta E(\mathbf{u})^{2}.\label{J}%
\end{equation}

\begin{lemma}
\label{GC}The functional defined by (\ref{J}) is $G$-compact (where $G$ is
defined by (\ref{ggg})).
\end{lemma}

\textbf{Proof. }Let $\mathbf{u}_{n}$ $=\left(  u_{n},\hat{u}_{n},\theta
_{n},\Theta_{n},\mathbf{E}_{n},\mathbf{H}_{n}\right)  $ be a minimizing
sequence for $J.$ Since the $G$-compactness depends on subsequences, we can
take a subsequence in which all the $C(\mathbf{u}_{n})$ have the same sign.
So, to fix the ideas, we can assume that
\[
C(\mathbf{u}_{n})>0;
\]
thus we have that
\[
J(\mathbf{u}_{n})=\frac{E(\mathbf{u}_{n})}{C(\mathbf{u}_{n})}+\delta
E(\mathbf{u}_{n})^{2}.
\]
It is immediate to see that $E(\mathbf{u}_{n})=\frac{1}{2}\left\Vert
\mathbf{u}_{n}\right\Vert ^{2}+\int N(u_{n})dx$ is bounded, moreover by
assumption (W-iiii)%
\begin{equation}
\left\vert \int_{Q}N(u_{n})dx\right\vert =o\left(  \left\Vert \mathbf{u}%
_{n}\right\Vert _{Q}^{2}\right)  .\label{pi}%
\end{equation}
We shall first show that
\begin{equation}
\left\Vert \mathbf{u}_{n}\right\Vert ^{2}\text{ is bounded.}\label{bibo}%
\end{equation}

Now $W\geq0$ and $E(\mathbf{u}_{n})$ is bounded. Then, comparing (\ref{placo})
with (\ref{nor}), in order to show that $\left\Vert \mathbf{u}_{n}\right\Vert
^{2}$ is bounded we have only to prove that
\begin{equation}
\left\Vert u_{n}\right\Vert _{L^{2}}\text{ is bounded.}\label{bouu}%
\end{equation}
Observe that, by using again the boundeness of $E(\mathbf{u}_{n})$, we have
that
\begin{equation}
\int W(u_{n})\text{ and }\int\left\vert \nabla u_{n}\right\vert ^{2}\text{ are
bounded.}\label{lim}%
\end{equation}

By (\ref{lim}) we have that
\begin{equation}
\int\left\vert u_{n}\right\vert ^{6}\text{ is bounded.}\label{limm}%
\end{equation}

Let $\varepsilon>0$ and set
\[
\Omega_{n}=\left\{  x\in\mathbb{R}^{3}:\left\vert u_{n}(x)\right\vert
>\varepsilon\right\}  \text{ and }\Omega_{n}^{c}=\mathbb{R}^{3}\backslash
\Omega_{n}.
\]
By (\ref{lim}) and since $W\geq0,$ we have
\begin{equation}
\int_{\text{ }\Omega_{n}^{c}}W(u_{n})\text{ is bounded .}\label{llim}%
\end{equation}
By (W-ii) we can write
\[
W(s)=\frac{m}{2}s^{2}+o(s^{2})\text{.}%
\]
Then, if $\varepsilon$ is small enough, there is a constant $c>0$ such that
\begin{equation}
\int_{\text{ }\Omega_{n}^{c}}W(u_{n})\geq c\int_{\Omega_{n}^{c}}u_{n}%
^{2}.\label{ma}%
\end{equation}
By (\ref{llim}) and (\ref{ma}) we get that
\begin{equation}
\int_{\Omega_{n}^{c}}u_{n}^{2}\text{ is bounded.}\label{po}%
\end{equation}

On the other hand
\begin{equation}
\int_{\Omega_{n}}u_{n}^{2}\leq\left(  \int_{\Omega_{n}}\left\vert
u_{n}\right\vert ^{6}\right)  ^{\frac{1}{3}}\cdot meas(\Omega_{n})^{\frac
{2}{3}}.\label{u}%
\end{equation}
By (\ref{limm}) we have that%
\begin{equation}
meas(\Omega_{n})\text{ is bounded.}\label{uu}%
\end{equation}
By (\ref{u}), (\ref{uu}), (\ref{limm}) we get that
\begin{equation}
\int_{\Omega_{n}}u_{n}^{2}\text{ is bounded.}\label{ff}%
\end{equation}
So (\ref{bouu}) follows from (\ref{po}) and (\ref{ff}).

We shall now prove that there exists \bigskip a subsequence $u_{n^{\prime}}$
of $u_{n}$ and
\begin{equation}
\left\{  j_{n^{\prime}}\right\}  \subset\mathbb{I=}\left\{  j\in\mathbb{Z}%
^{3}:C_{Q_{j}}\left(  u_{n}\right)  >0\right\} \label{ba}%
\end{equation}
such that%
\begin{equation}
g_{j_{n^{\prime}}}u_{n^{\prime}}\rightharpoonup\bar{u}\text{ }\neq0\text{
weakly in }H^{1}\left(  \mathbb{R}^{3}\right)  .\label{bi}%
\end{equation}

To this end we show first that for any $n$ sufficiently large there is
$j_{n}\in\mathbb{I}$ such that (see (\ref{kkk}))
\begin{equation}
\frac{E_{Q_{j_{n}}}(\mathbf{u}_{n})}{C_{Q_{j_{n}}}(\mathbf{u}_{n})}\leq
\Lambda_{0}-\frac{b}{2}.\label{first}%
\end{equation}

Since $\mathbf{u}_{n}$ is a minimizing sequence for $J$, by (\ref{kkk}) there
is $M>0$ such that, for any $n\geq M$ we have%

\begin{align}
\Lambda_{0}-\frac{b}{2}  & \geq J(\mathbf{u}_{n})=\frac{E(\mathbf{u}_{n}%
)}{C(\mathbf{u}_{n})}+\delta E(\mathbf{u}_{n})^{2}\nonumber\\
& \geq\frac{E(\mathbf{u}_{n})}{C(\mathbf{u}_{n})}=\frac{\sum_{j}E_{Q_{j}%
}(\mathbf{u}_{n})}{\sum_{j}C_{Q_{j}}(\mathbf{u}_{n})}\nonumber\\
& \geq\frac{\sum_{j\in\mathbb{I}}E_{Q_{j}}(\mathbf{u}_{n})}{\sum
_{j\in\mathbb{I}}C_{Q_{j}}(\mathbf{u}_{n})}.\label{int}%
\end{align}
Now arguing by contradiction assume that (\ref{first}) does not hold, namely
assume that%
\begin{equation}
\text{for any }j\in\mathbb{I}\text{: }\frac{E_{Q_{j}}(\mathbf{u}_{n}%
)}{C_{Q_{j}}(\mathbf{u}_{n})}>\Lambda_{0}-\frac{b}{2}.\label{contr}%
\end{equation}
Then by (\ref{int}) and (\ref{contr}) we have
\begin{equation}
\Lambda_{0}-\frac{b}{2}\geq\frac{\sum_{j\in\mathbb{I}}E_{Q_{j}}(\mathbf{u}%
_{n})}{\sum_{j\in\mathbb{I}}C_{Q_{j}}(\mathbf{u}_{n})}>\frac{\sum
_{j\in\mathbb{I}}C_{Q_{j}}(\mathbf{u}_{n})\left(  \Lambda_{0}-\frac{b}%
{2}\right)  }{\sum_{j\in\mathbb{I}}C_{Q_{j}}(\mathbf{u}_{n})}=\Lambda
_{0}-\frac{b}{2}.\label{tree}%
\end{equation}

So we get a contradiction and then (\ref{first}) is proved.

Clearly $g_{j_{n}}\mathbf{u}_{n}$ is bounded, then there is a subsequence
$g_{j_{n^{\prime}}}\mathbf{u}_{n^{\prime}}$ such that
\begin{equation}
g_{j_{n^{\prime}}}u_{n^{\prime}}\rightharpoonup\bar{u}\text{ weakly in }%
H^{1}\left(  \mathbb{R}^{3}\right)  ,\text{ }g_{j_{n^{\prime}}}\theta
_{n^{\prime}}\rightharpoonup\bar{\theta}\text{ weakly in }L^{2}\left(
\mathbb{R}^{3}\right)  .\label{ueaa}%
\end{equation}
Here clearly we have set $\left(  g_{j_{n^{\prime}}}u_{n^{\prime}}\right)
(x)=u_{n^{\prime}}(x-j_{n^{\prime}})$ and $\left(  g_{j_{n^{\prime}}}%
\theta_{n^{\prime}}\right)  (x)=\theta_{n^{\prime}}(x-j_{n^{\prime}}). $

We show that
\begin{equation}
\left\Vert g_{j_{n^{\prime}}}\mathbf{u}_{n^{\prime}}\right\Vert _{Q}\text{
does not converge to }0.\label{inf}%
\end{equation}
Arguing by contradiction assume that
\begin{equation}
\left\Vert g_{j_{n^{\prime}}}\mathbf{u}_{n^{\prime}}\right\Vert _{Q}%
\rightarrow0.\label{infi}%
\end{equation}
Then by (\ref{first}) and (\ref{pi}) we have%
\begin{equation}
\Lambda_{0}-\frac{b}{2}\geq\frac{E_{Q_{j_{n^{\prime}}}}(\mathbf{u}_{n^{\prime
}})}{C_{Q_{j_{n^{\prime}}}}(\mathbf{u}_{n^{\prime}})}\geq\frac{\frac{1}%
{2}\left\Vert g_{j_{n^{\prime}}}u_{n^{\prime}}\right\Vert _{Q}^{2}%
+o(\left\Vert g_{j_{n^{\prime}}}u_{n^{\prime}}\right\Vert _{Q}^{2})}%
{C_{Q}(g_{j_{n^{\prime}}}\mathbf{u}_{n^{\prime}})}.\label{ca}%
\end{equation}
So by (\ref{infi}), by definition of $\Lambda_{0}$ and passing to the limit in
(\ref{ca}), we get%
\[
\Lambda_{0}-\frac{b}{2}\geq\Lambda_{0}%
\]
which gives the contradiction and (\ref{inf}) holds.

By (\ref{inf}) and (\ref{ueaa}) we deduce that
\[
\bar{u}\neq0\text{ in }Q.
\]
In fact, arguing by contradiction, assume that $\bar{u}=0$ in $Q.$ By
(\ref{ueaa}) we get $g_{j_{n^{\prime}}}u_{n^{\prime}}\rightarrow\bar{u}=0$
strongly in $L^{2}(Q)$ and $g_{j_{n^{\prime}}}\theta_{n^{\prime}%
}\rightharpoonup\bar{\theta}$ weakly in $L^{2}\left(  Q\right)  ,$ then
\begin{equation}
C_{Q}(g_{j_{n^{\prime}}}\mathbf{u}_{n^{\prime}})=\int_{Q}g_{j_{n^{\prime}}%
}u_{n^{\prime}}g_{j_{n^{\prime}}}\theta_{n^{\prime}}\rightarrow0\label{bene}%
\end{equation}
By (\ref{ca}) and (\ref{bene}) we deduce that
\begin{equation}
E_{Q}(g_{j_{n^{\prime}}}\mathbf{u}_{n^{\prime}})\rightarrow0\label{issimo}%
\end{equation}
and consequently $\left\Vert g_{j_{n^{\prime}}}\mathbf{u}_{n^{\prime}%
}\right\Vert _{Q}\rightarrow0$ contradicting (\ref{inf}). So we conclude that
$\bar{u}\neq0$ in $Q.$

From now on we write for simplicity $g_{j_{n}}\mathbf{u}_{n}$ instead of
$g_{j_{n^{\prime}}}\mathbf{u}_{n^{\prime}}$ and set%

\[
g_{j_{n}}\mathbf{u}_{n}=\mathbf{\bar{u}+\mathbf{w}_{n}.}%
\]
with $\mathbf{\mathbf{w}_{n}}\rightharpoonup0$ weakly.

We finally show that there is no splitting, namely that $\mathbf{\mathbf{w}%
_{n}\rightarrow0}$ strongly.

By the $G$-invariance of $E$ and $C$ and lemma \ref{tre}, we have%
\begin{align*}
J_{\ast}  & :=\lim J(g_{j_{n}}\mathbf{u}_{n})=\lim\frac{E(g_{j_{n}}%
\mathbf{u}_{n})}{C(g_{j_{n}}\mathbf{u}_{n})}+\delta E(g_{j_{n}}\mathbf{u}%
_{n})^{2}\\
& =\lim\frac{E(\mathbf{\bar{u}})+E(\mathbf{w}_{n})+o(1)}{C(\mathbf{\bar{u}%
})+C(\mathbf{w}_{n})+o(1)}+\delta\left[  E(\mathbf{\bar{u}})+E(\mathbf{w}%
_{n})+o(1)\right]  ^{2}\\
& =\lim\frac{E(\mathbf{\bar{u}})+E(\mathbf{w}_{n})}{C(\mathbf{\bar{u}%
})+C(\mathbf{w}_{n})}+\delta E(\mathbf{\bar{u}})^{2}+\delta E(\mathbf{w}%
_{n})^{2}+2\delta E(\mathbf{\bar{u}})E(\mathbf{w}_{n})\\
& \geq\lim\frac{E(\mathbf{\bar{u}})+E(\mathbf{w}_{n})}{\left\vert
C(\mathbf{\bar{u}})\right\vert +\left\vert C(\mathbf{w}_{n})\right\vert
}+\delta E(\mathbf{\bar{u}})^{2}+\delta E(\mathbf{w}_{n})^{2}+2\delta
E(\mathbf{\bar{u}})E(\mathbf{w}_{n})\\
& \geq\lim\left[  \min\left(  \frac{E(\mathbf{\bar{u}})}{\left\vert
C(\mathbf{\bar{u}})\right\vert },\frac{E(\mathbf{w}_{n})}{\left\vert
C(\mathbf{w}_{n})\right\vert }\right)  \right]  +\delta E(\mathbf{\bar{u}%
})^{2}+\delta E(\mathbf{w}_{n})^{2}+2\delta E(\mathbf{\bar{u}})E(\mathbf{w}%
_{n})\\
& \geq\lim\left[  \min\left(  J(\mathbf{\bar{u}}),J(\mathbf{w}_{n})\right)
\right]  +2\delta E(\mathbf{\bar{u}})E(\mathbf{w}_{n})\geq(\text{since
}J_{\ast}=\inf J)\\
& \geq\lim\left[  J_{\ast}+2\delta E(\mathbf{\bar{u}})E(\mathbf{w}%
_{n})\right]  =J_{\ast}+2\delta E(\mathbf{\bar{u}})\lim E(\mathbf{w}_{n}).
\end{align*}
Then%
\[
2\delta E(\mathbf{\bar{u}})\lim E(\mathbf{w}_{n})\leq0
\]
and since $E(\mathbf{\bar{u}})\neq0,$ we have that%
\[
\lim E(\mathbf{w}_{n})=0
\]
and hence $\mathbf{u}_{n}\rightarrow\mathbf{\bar{u}}$ strongly. Then $J$ is
$G$-compact.

$\square$\bigskip

\textbf{Proof of Th. \ref{main-theorem}}. We shall use Theorem \ref{astra}.
Obviously assumptions (EC-1) and (EC-2) are satisfied with $G$ given by
(\ref{ggg}). Then by lemma \ref{GC} and Th. \ref{astra}, we have the existence
of soliton solutions. In order to prove that they form a family dependent of
$\delta$, it is sufficient to prove that $\delta_{1}\neq\delta_{2}$ in the
definition (\ref{J}) of $J$ implies $\mathbf{u}_{\delta_{1}}\neq
g\mathbf{u}_{\delta_{2}}$ for every $g\in G.$ We argue indirectly and assume
that $\mathbf{u}_{\delta_{1}}=g\mathbf{u}_{\delta_{2}}$ for some $g\in G.$
Then%
\[
\frac{E(g\mathbf{u}_{\delta_{2}})}{\left\vert C(g\mathbf{u}_{\delta_{2}%
})\right\vert }+\delta_{2}E(g\mathbf{u}_{\delta_{2}})^{2}=\frac{E(\mathbf{u}%
_{\delta_{1}})}{\left\vert C(\mathbf{u}_{\delta_{1}})\right\vert }+\delta
_{1}E(\mathbf{u}_{\delta1})^{2}%
\]
and so, since $g\mathbf{u}_{\delta_{2}}=\mathbf{u}_{\delta_{1},}$
\begin{align*}
0  & =\frac{E(g\mathbf{u}_{\delta_{2}})}{\left\vert C(g\mathbf{u}_{\delta_{2}%
})\right\vert }+\delta_{2}E(g\mathbf{u}_{\delta_{2}})^{2}-\left(
\frac{E(\mathbf{u}_{\delta_{1}})}{\left\vert C(\mathbf{u}_{\delta_{1}%
})\right\vert }+\delta_{1}E(\mathbf{u}_{\delta1})^{2}\right) \\
& =\left(  \delta_{2}-\delta_{1}\right)  E(\mathbf{u}_{\delta_{1}})^{2}%
\end{align*}
Then, since $\delta_{1}\neq\delta_{2},$ $E(\mathbf{u}_{\delta_{1}})=0$ and so
$\mathbf{u}_{\delta_{1}}=0$ which is a contradiction.

$\square$

\bigskip

\textbf{Proof of Th. \ref{imp}. }Let $\mathbf{u}_{\delta}=\left(  u_{\delta
},0,\theta_{\delta},0,\mathbf{E}_{\delta},\mathbf{0}\right)  $ be a minimizer
of $J$ (defined by (\ref{J})) on $X=\left\{  \mathbf{u}\in\mathcal{H}%
:\nabla\cdot\mathbf{E}=-q\theta u,\ \nabla\cdot\mathbf{H}=0\right\}  .$ By
Theorem\ref{astraco} any $\mathbf{u}\in X,$ with energy $E(\mathbf{u}_{\delta
})$ and charge $C(\mathbf{u}_{\delta}),$ is a soliton. So, in particular also
$\mathbf{u}_{\delta}$ is a soliton. Clearly $\mathbf{u}_{\delta}$ minimizes
also the energy $E$ (see (\ref{placo})) on the manifold%
\[
X_{\delta}=\left\{  \mathbf{u}\in X:C(\mathbf{u)}=C(\mathbf{u}_{\delta
})=\sigma_{\delta}\right\}  .
\]
If we write $\mathbf{E}=-\nabla\phi$, the constraint $\nabla\cdot
\mathbf{E}=-q\theta u$ becomes
\[
\Delta\phi=q\theta u.
\]
So $\mathbf{u}_{\delta}$ is a critical point of $E$ on the manifold (in
$\mathcal{H)}$ made up by those $\mathbf{u}=$ $\left(  u,0,\theta,0,\nabla
\phi,\mathbf{0}\right)  $ satisfying the constraints%
\begin{equation}
\Delta\phi=q\theta u\label{vic1}%
\end{equation}

\begin{equation}
C(\mathbf{u)=}\int\theta u\ dx=\sigma_{\delta}.\label{vic2}%
\end{equation}

Therefore, for suitable Lagrange multipliers $\lambda\in\mathbb{R},$ $\xi
\in\mathcal{D}^{1,2}$ ($\mathcal{D}^{1,2}$ is the closure of $C_{0}^{\infty}$
with respect to the norm $\left\Vert \nabla\phi\right\Vert _{L^{2}}$), we have
that $\mathbf{u}_{\delta}$ is a critical point of
\begin{equation}
E_{\lambda,\xi}(\mathbf{u)}=E(\mathbf{u)+}\lambda\left(  \int\theta
u\ dx\mathbf{-}\sigma_{\delta}\right)  +\left\langle \xi,-\Delta\phi+q\theta
u\right\rangle \label{lag}%
\end{equation}
where $\left\langle \ \cdot\ ,\ \cdot\ \right\rangle $ denotes the duality map
in $\mathcal{D}^{1,2}.$ It is easy to show that $E_{\lambda,\xi}^{\prime
}(\mathbf{u}_{\delta})=0$ gives the equations%
\begin{align}
-\Delta u_{\delta}+W^{\prime}(u_{\delta})+\lambda\theta_{\delta}+q\xi
\theta_{\delta}  & =0\label{var1}\\
-\Delta\phi_{\delta}  & =\Delta\xi\label{var2}\\
\theta_{\delta}+\lambda u_{\delta}+q\xi u_{\delta}  & =0.\label{var3}%
\end{align}
From (\ref{var2}) we get $\xi=-\phi_{\delta},$ so (\ref{var1}) and
(\ref{var3}) become%
\begin{align*}
-\Delta u_{\delta}+W^{\prime}(u_{\delta})+\theta_{\delta}(\lambda
-q\phi_{\delta})  & =0\\
\left(  \lambda-q\phi_{\delta}\right)  u_{\delta}  & =-\theta_{\delta}.
\end{align*}
From the above equations we clearly get (\ref{stat1}). (\ref{stat2}) is given
by the constraint (\ref{vic1}).

$\square$

\end{document}